\documentclass{raa}            

\newcommand\natcom{{\it Nat.~Commun.\ }}
\newcommand\sces{{\it Sci.~China~Earth~Sci.\ }}

\usepackage{natbib}

\newcommand{\app}{    {\it Astropart. Phys.}}

\newcommand{\rpp}{    {\it Rep. Progr. Phys.}}
\newcommand{\sola}{   {\it Solar Phys.}}

\usepackage{multirow}
\usepackage{ulem}
\usepackage{graphicx,times}             

\begin{document}

   \title{{Energy Spectral property in an Isolated CME-driven Shock}
   \footnote{This work is supported by the Xinjiang Natural Science
Foundation No. 2014211A069}
}

   \volnopage{Vol.0 (200x) No.0, 000--000}      
   \setcounter{page}{1}          

   \author{Xin Wang
      \inst{1,2,3,4}, Yihua Yan\inst{2}, Mingde Ding\inst{3}, Na Wang\inst{1,4},\and Hao Shan\inst{1,4}
   }
\institute{Xinjiang Astronomical Observatory, Chinese Academy of
Sciences, Urumqi 830011, China; {\it e-mail: wangxin@xao.ac.cn}\\
{Key Laboratory of Solar Activities, National Astronomical
Observatories, Chinese Academy of Sciences, Beijing 100012, China}
\\{Key Laboratory of Modern Astronomy and Astrophysics (Nanjing
University)£¬Ministry of Education, Nanjing 210093, China}\\{Key
Laboratory of Radio Astronomy, Chinese Academy of Sciences, Nanjing
210008, China}}



\abstract{Observations from multiple spacecraft show that there are
energy spectral ``breaks" at 1-10MeV in some large CME-driven
shocks. However, numerical models can hardly simulate this property
due to high computational expense. The present paper focuses on
analyzing these energy spectral ``breaks" by Monte Carlo particle
simulations of an isolated CME-driven shock. Taking the Dec 14 2006
CME-driven shock as an example, we investigate the formation of this
energy spectral property. For this purpose, we apply different
values for the scattering time in our isolated shock model to obtain
the highest energy ``tails", which can potentially exceed the
``break" energy range.  However, we have not found the highest
energy ``tails" beyond the ``break" energy range, but instead find
that the highest energy ``tails" reach saturation near the range of
energy at 5MeV. So, we believe that there exists an energy spectral
``cut off" in an isolated shock. If there is no interaction with
another shock, there would not be formation of the energy spectral
``break" property. }

   \authorrunning{X. Wang \& et al}            
   \titlerunning{ENERGY SPECTRAL PROPERTY IN A CME-DRIVEN SHOCK}  

   \maketitle

%
%
\section{Introduction}           
\label{sect:intro}

Strong astrophysical shocks are often associated with superthermal particle emission and with magnetic field
amplification \citep{bmr13,veb06}. This phenomenon suggests that shocks are regions where particles are efficiently
accelerated, and this large group of energetic particles is responsible for the excitation of magnetic turbulence via
plasma instabilities \citep{bell78,bsr13,Jokipii13}. These magnetic fields which diffuse cosmic rays in the vicinity of
the shock are required to be much higher than the averaged magnetic field in the interstellar medium.

The theoretical model includes the determination of the particle injection energy from the thermal particle
distribution into the non-thermal particle distribution, the maximum energy of particles accelerated at the shock,
energetic particle spectra at all spatial and temporal locations, and the dynamical distribution of particles that
escape upstream and downstream from the evolving shock complex \citep{zrw00}. Monte Carlo simulation results indicate
that solar ejecta transfer energy into the non-thermal particles in an interplanetary shock with an efficiency of
$\sim$ 10\% \citep{wwy13}. Studies of the dependence of this efficiency on the angle between shock normal and the
magnetic field direction ($\theta_{BN}$)can have implications for ground-level enhancement (GLE) events
\citep{llc10,sro13}. Estimation of the maximum particle energy by CME-driven shocks is becoming more and more vital for
forecasting space weather. Since particles accelerated at the shock escape rather easily from the acceleration site,
they can be detected well before the arrival of the shock. This, of course, has immediate and interesting implications
for space weather monitoring and prediction systems, but it does imply too that the study of the ion acceleration
mechanism is complicated by the subsequent interplanetary propagation of the energetic particles.

For the past several decades, there has been much literature focusing on all aspects of the diffusive shock
acceleration (DSA). In the past, cosmic ray (CR) spectra, acceleration efficiency, and amplification of the magnetic
field have been calculated regularly via a two-fluid approach \citep{drury83}. More recently, those have been simulated
via a particle Monte Carlo method \citep{veb06,wy11,ed04,emp90,no04}, or via hybrid method
\citep{cs14,gs12,gbs93,gg13,Winske85}, and via full particle-in-cell (PIC) method \citep{ah07,rs11}. In addition, the
CR's transport equations have also been solved by a numerical method \citep{kjg02,za10} and an analytical method
\citep{liu04,cab10,mv96}. These methods are all able to provide consistent results for the dynamics of the shock
including the CR's back-reaction. However, unlike the analytical method, the particle method and the numerical MHD
method have not yet been able to simulate the energy spectral ``break" property \citep{md13}. Since the ``break" of the
energy spectrum would be associated with the particle leakage mechanism, Malkov has presented a new combined diffusion
coefficient to describe particle acceleration and escape in different regions. It accounts for a high turbulent
magnetic field in the vicinity of the shock site (particle acceleration) and for faded turbulence of the magnetic field
far away from the shock front (particle escape).

Although it is widely accepted that the most efficient acceleration of solar energetic particles(SEP) would happen in
CME-driven shocks, the underlying acceleration mechanism in the shock environment still remains uncertain. In
particular, it is not clear how the extensive maximum particle energy can be produced and why the energy spectral shape
can be broken (i.e., why an abrupt change in slope of the energy spectrum can occur ) in some large CME-driven shocks
\citep{mewaldt08}. In the past solar cycle 23, there are several observed events exhibiting proton energy spectral
``breaks". These events occurred on Nov-6-97, Apr-15-01, Jan-20-05, Sep-7-05, Dec-05-06, and Dec-14-06, respectively.
In addition, there are hard X-ray and $\gamma$-ray energy spectra from Reuven Ramaty High Energy Solar Spectroscopic
Imager (RHESSI) occurring on 2002 July 23. This event shows a double-power-law spectrum with a ``break" at $\sim$30keV
in X-ray and a high energy ``cut-off" tail at $\sim$5MeV in $\gamma$-ray. The X-ray spectrum indicates that the
substantial electron acceleration achieved to tens of keV. The $\gamma$-ray line shows that ions are accelerated to
tens of MeV\citep{lkh03}. There are also some debates about broken lower energy spectrum in X-ray, which is far
different from an ad hoc assumption of hot thermal plasma presenting as the highest low-energy cutoff ($\sim$20keV).
Actually, there are a lot of analyses of the hardening spectra occurring in the energy range varying from 20keV to a
few MeV \citep{glc01,klc13,huang09}. More recent years, an extensive solar energetic particle (SEP) event was detected
by STEREO A on July 23 2012 near 1 AU. \citet{ll14} suggest that the in-transit interaction between two closely
launched CMEs resulted in the extreme enhancement of the SEP. These results provide a new view crucial to space weather
and solar physics as to how an extreme space weather event can be produced from an interaction between solar energetic
ejecta \citep{Gopal05,czd13,wj13,svh13,schneider93}.

The parallel shocks show an effective amplification of the initial magnetic field due to the current of energetic ions
that propagate anisotropically into the upstream flow. \citet{cs13,cs14} have used 2D and 3D hybrid simulations with
large computational boxes to reveal the formation of upstream filaments and cavities, which eventually trigger the
Richtmeyer--Meshkov instability at the shock, and lead to further turbulent amplification of magnetic fields in the
downstream region. The typical acceleration time, up to energy $E$ in a shock with its velocity $v_{sh}$, is of order
$T_{acc}\approx D(E)/v_{sh}^2$ \citep{drury83}, where $D(E)$ is the diffusion coefficient. The acceleration
characteristic time scale would be roughly equivalent to the time scale of the ejecta-dominated stage: when the shock
is formed, the shock velocity drops quite rapidly, and so does amplification of the magnetic field \citep{gg12}. In a
certain acceleration time scale of the shock system, the maximum particle energy is decided. However, in the particle
simulation system, the particle's free escaped boundary (FEB) size should be considered, that means the highest energy
particle would be escaped from the FEB. If the present simulation will focus on production of the maximum particle
energy, the highest energy spectral ``tail" should be reserved. To obtain the maximum particle energy, we can either
add the FEB size or decrease the value for the scattering time. Due to the expansion of the FEB size, the shock system
will be brought the extra computational burden, so we can change the value for the scattering time. In the amplified
magnetic field with an order of magnitude $\delta{B}/B_{0}\sim 1$, the scattering time is an important factor to
determine the acceleration efficiency in the resonant diffusion condition, and thereby determine if the maximum
particle energy can be achieved. In this isolated shock model, we can investigate the maximum particle energy by
changing the value for the scattering time.

Because we are not sure if an isolated CME-driven shock can accelerate energetic particles beyond E$_{break}$ and even
up to GeV, we take an isolated shock as an example to investigate the maximum particle energy and energy spectral
``break"  by using different values for the scattering time within resonant diffusive scenarios. According to the DSA
theory, acceleration efficiency is significantly enhanced once the mean free path for pitch-angle scattering is
approximately equal to the particle gyroradius (i.e. $\lambda\approx r_L(E)\propto E/B$), and the diffusion coefficient
reads $D_B(E)\approx v r_L(E)$ \citep{lc83}. If the Bohm diffusion condition is satisfied in the shock system and a
typical interplanetary magnetic field with an order of a few mG, one can estimate that the maximum particle energy
would be $E_{max}\approx 1$ MeV, which is not enough to explain the energy spectral ``break" at 1-10MeV in
observations\citep{emp90}. Therefore, we hope to extensively calculate the maximum particle energy E$_{max}$ using
different values for the scattering time within an isolated shock model. If we can obtain the E$_{max}$ $>$
E$_{break}$, it would imply that there is unnecessary to use multiple shocks model to explain the energy spectral
``break" property. If we obtain the E$_{max}$ $<$ E$_{break}$, then we should examine that whether there would exist an
energy spectral ``break" at E$_{break}$ and whether we should need a multiple shocks model indeed.

In present paper, we do simulations to further investigate the maximum particle energy in an isolated CME shock by
using different values for the scattering time. In section 2, we introduce the dynamical Monte Carlo simulation method
simply. In section 3, the simulated results and analysis are presented. In the end, we give a summary and some
conclusions.


\section{Method}
\label{Method} Many deviations of DSA arisen from the nonlinear effects of the shock, such as the modification of the
shock structure, magnetic field obliquity, time-dependence, magnetic field amplification, and etc. Those have been
calculated by a two-fluid model \citep{dv81} and an analytical model \citep{cab10,mv96,ab06} and particle models
including hybrid, particle-in-cell,and Monte-carlo method \citep{gs12,Giacalone04,ah07,rs11,veb06,ed04,wy11}. These
models return consistent results well, also can provide the results of the dynamics of the shock including the CR's
back-reactions. In general, there are two aspects of the deviation of DSA: one aspect is about the transfer issue
depending on the macro factors of the shock in Mach number, magnetic field obliquity, and time-dependence etc; another
aspect is about the acceleration issue depending on micro factors of the shock in diffusive coefficient, injection
rate, and scattering time etc.

Here, we use a dynamical Monte Carlo method to study the acceleration issue depending on factor of scattering time. In
this isolated shock model, the maximum particle energy $E_{max}$ will be calculated in different cases by applying
different values for the constant of the scattering time. Since the FEB measures the size of the faded turbulent
magnetic field in shock precursor region, if the FEB size is larger, then the $E_{max}$ is higher. Unfortunately, if
the size of the FEB is larger, then the computational expense is higher. Instead, we can change the scattering time to
achieve a higher $E_{max}$ in the shock. Assuming a particle can obtain the same additional energy gain from each cycle
in a period of the scattering time, it is probable that the more scattering probabilities will obtain more energy
gains. If we take a smaller value for the constant of the scattering time in one simulation case, we can obtain a
higher $E_{max}$ value by more scattering probabilities in total simulation time.

Although there still exists the impact from the diffusive coefficient in different shock region, it can be neglected in
this isolated shock model. Since the ejecta moves with a large speed and the most part of accelerated particles are
located in the turbulent magnetic field in the vicinity of ejecta, so the diffusive processes always can be taken as
the Bohm condition and its coefficient difference in this limited precursor region would be slight.

Monte Carlo approach regards the fluid as being composed of particles and focuses on the scattering micro processes
between the particles and turbulent magnetic field in the diffusion processes. This technique is based on the
computational grids, where large number of particles distributed. Particle's mean free path is proportional to its
local velocity in its local frame as follows.
\begin{equation}\label{equ1}
    \lambda=\upsilon_{L}\cdot\tau
\end{equation}
where, $\upsilon_{L}$ is the local velocity of particles, $\tau$ is the scattering time. In  Earth's bow shock model,
the scattering time $\tau$ is taken as a constant\citep{kje96}. For comparing values for maximum particle energy
$E_{max}$, we apply different values for constant of the scattering time to perform these corresponding cases as in
Table \ref{tab:para}. To simulate the scattering processes accurately, the scattering time $\tau$ should be chosen to
be far more than time step $dt$ as follows.
\begin{equation}\label{equ2}
    \tau \gg dt
\end{equation}

To simulate the shock's formation and evolution, we set the standard scattering time $\tau_{0}$ is a constant for all
particles in Case A. Other constants of the scattering time and corresponding cases can be seen in Table\ref{tab:para}.
Those related simulation parameters can be refereed to previous work \citep{wy12}. Here, we just list the scattering
times in different cases. All of the scattering times are chosen to be more than time step dt ($dt$=$\tau_{0}$/25) in
corresponding cases.




\begin{table} 
\begin{center}
 \caption{\label{tab:para} Six cases with corresponding constants of the
 scattering time}
  \medskip\medskip
  \begin{tabular}{|l|c|c|c|c|c|c|}
  \hline 
  Simulation cases & A & B & C & D & E &F \\
  \hline
  The scattering  time & $\tau_{0}$ & $\tau_{0}$/2 & $\tau_{0}$/3 & $\tau_{0}$/4 & $\tau_{0}$/5 & $\tau_{0}$/12.5 \\
  \hline
\end{tabular}
\end{center}
\end{table}




\section{Results}
\subsection{Acceleration of particles}
 \begin{figure}[ht]
\centering
  \includegraphics[width=2.8in, angle=0]{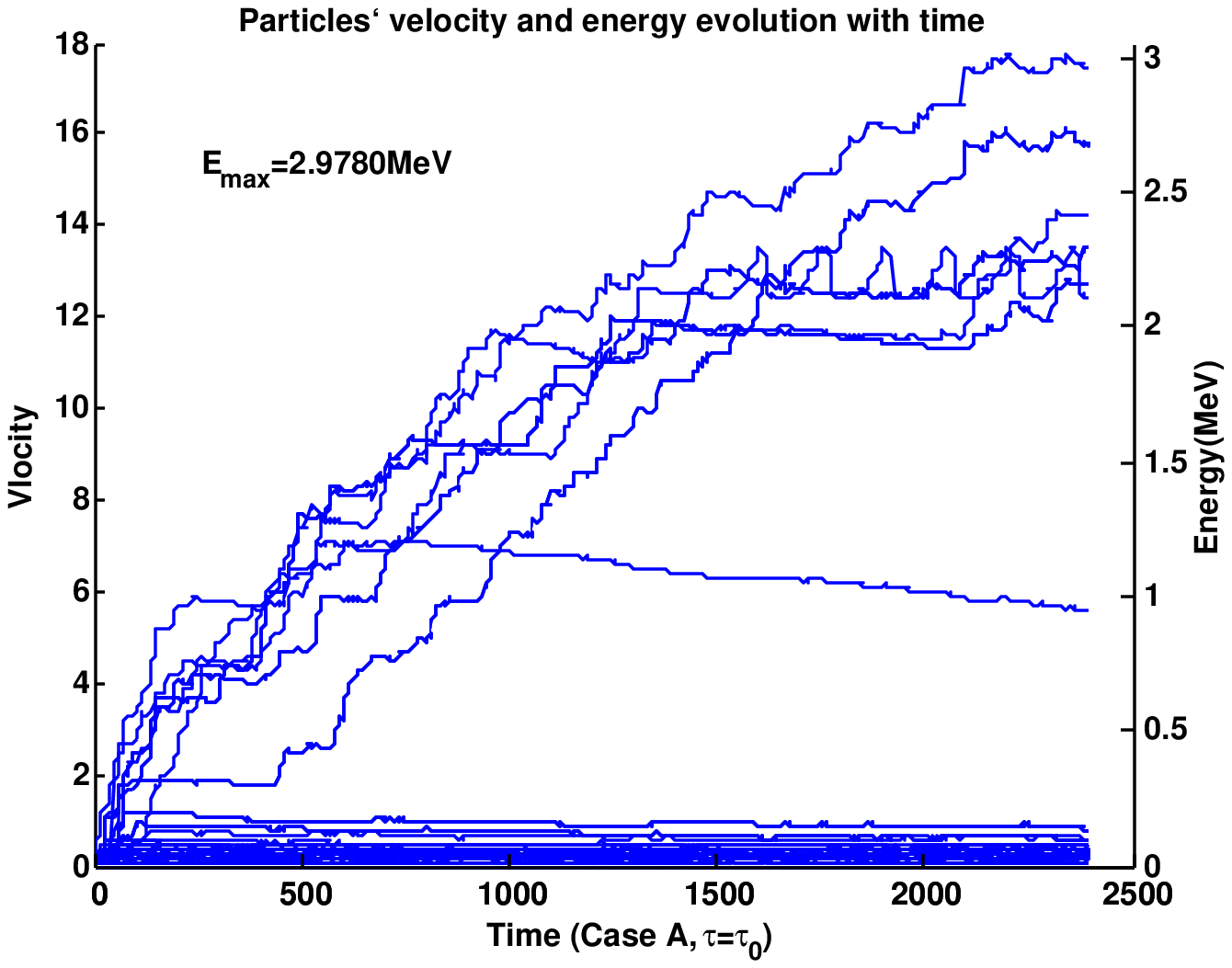}
  \includegraphics[width=2.8in, angle=0]{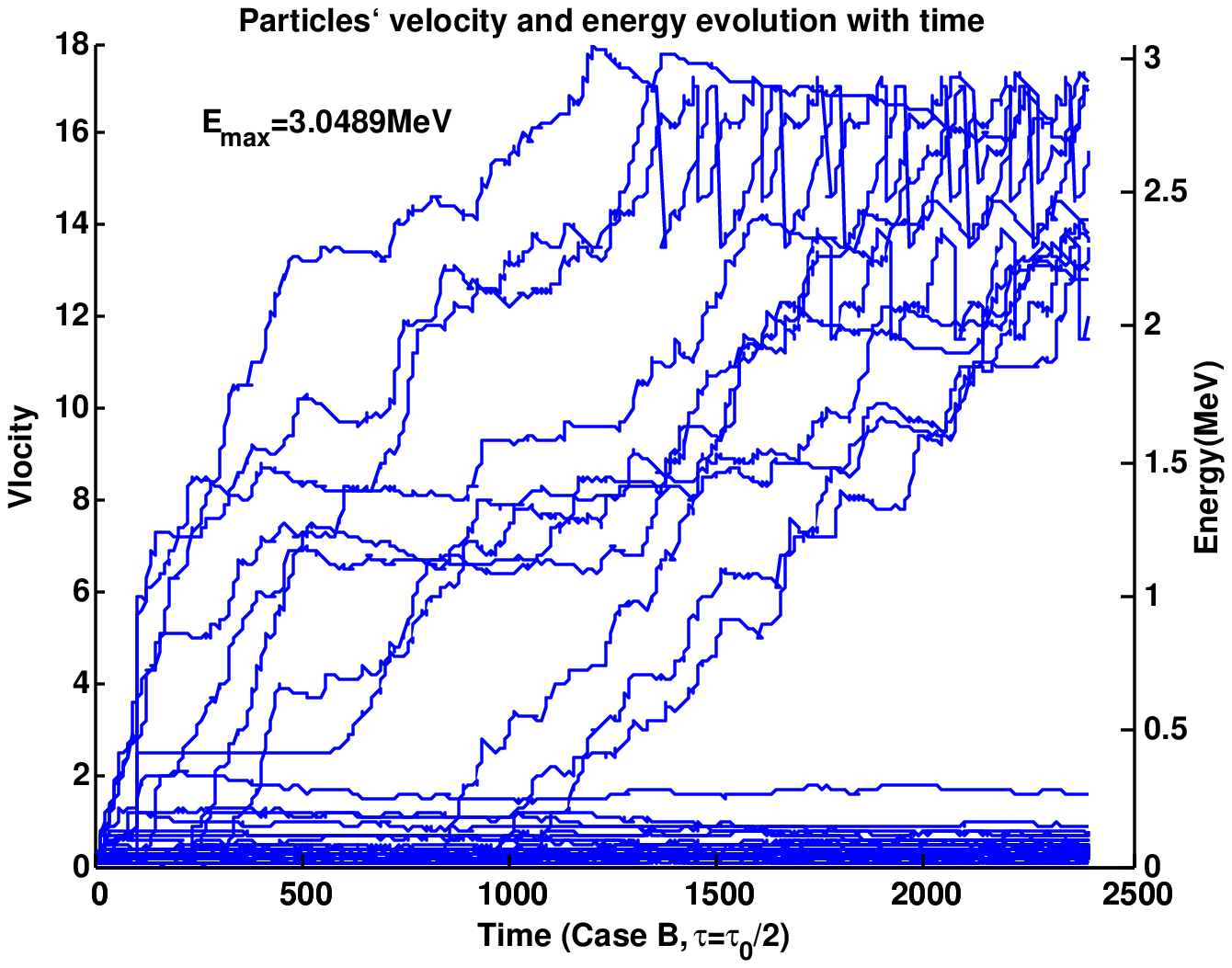}\\
  \includegraphics[width=2.8in, angle=0]{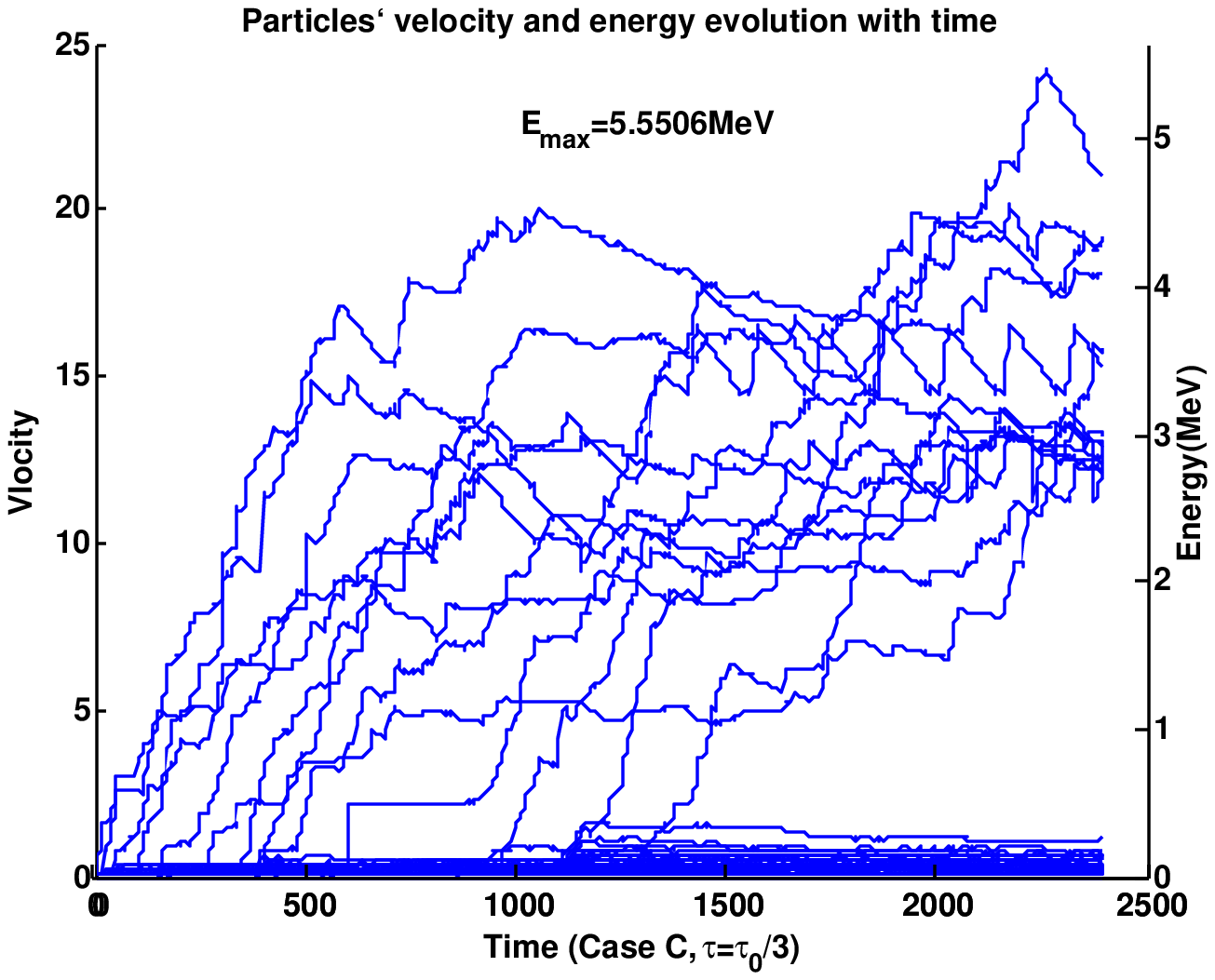}
  \includegraphics[width=2.8in, angle=0]{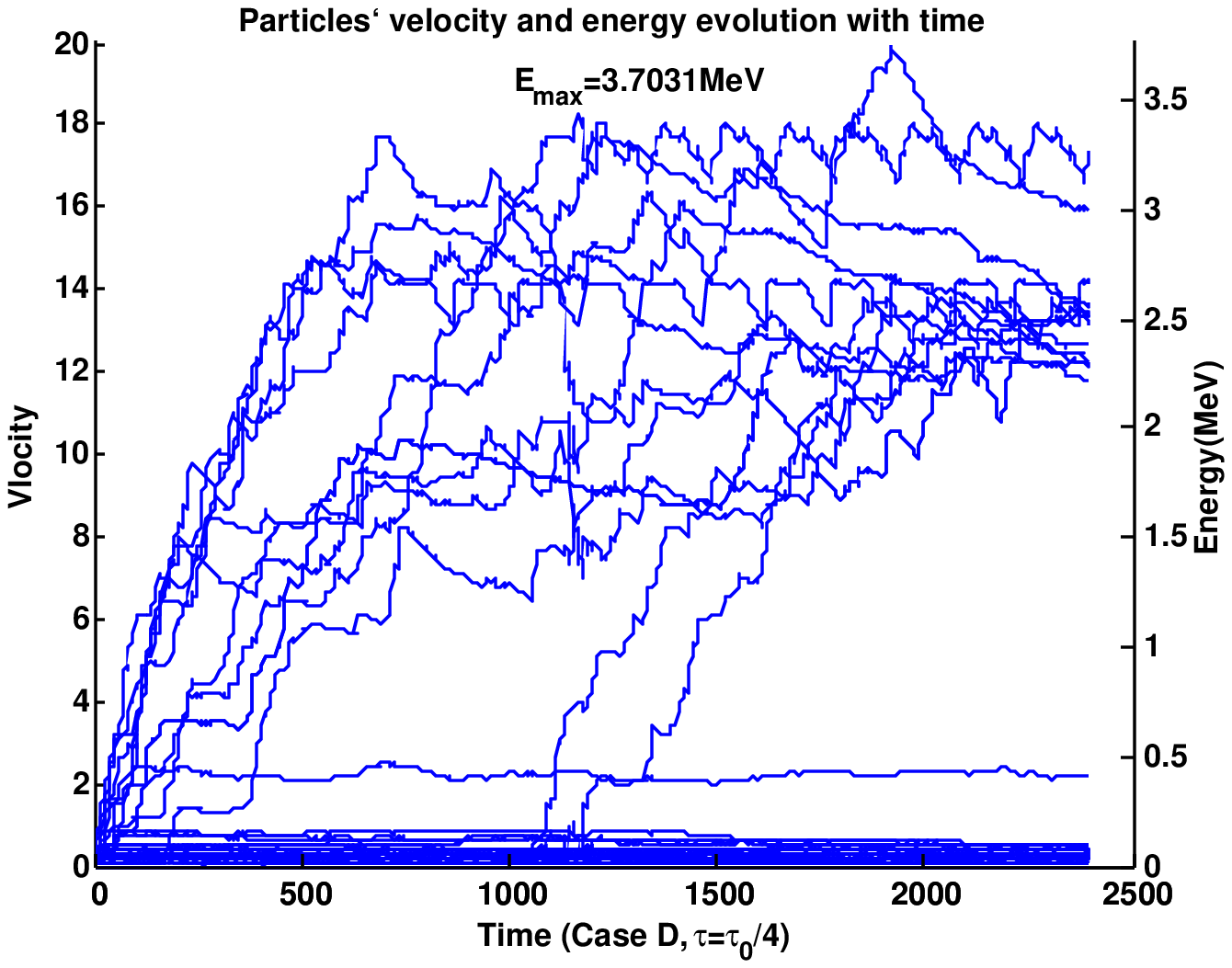}\\
  \includegraphics[width=2.8in, angle=0]{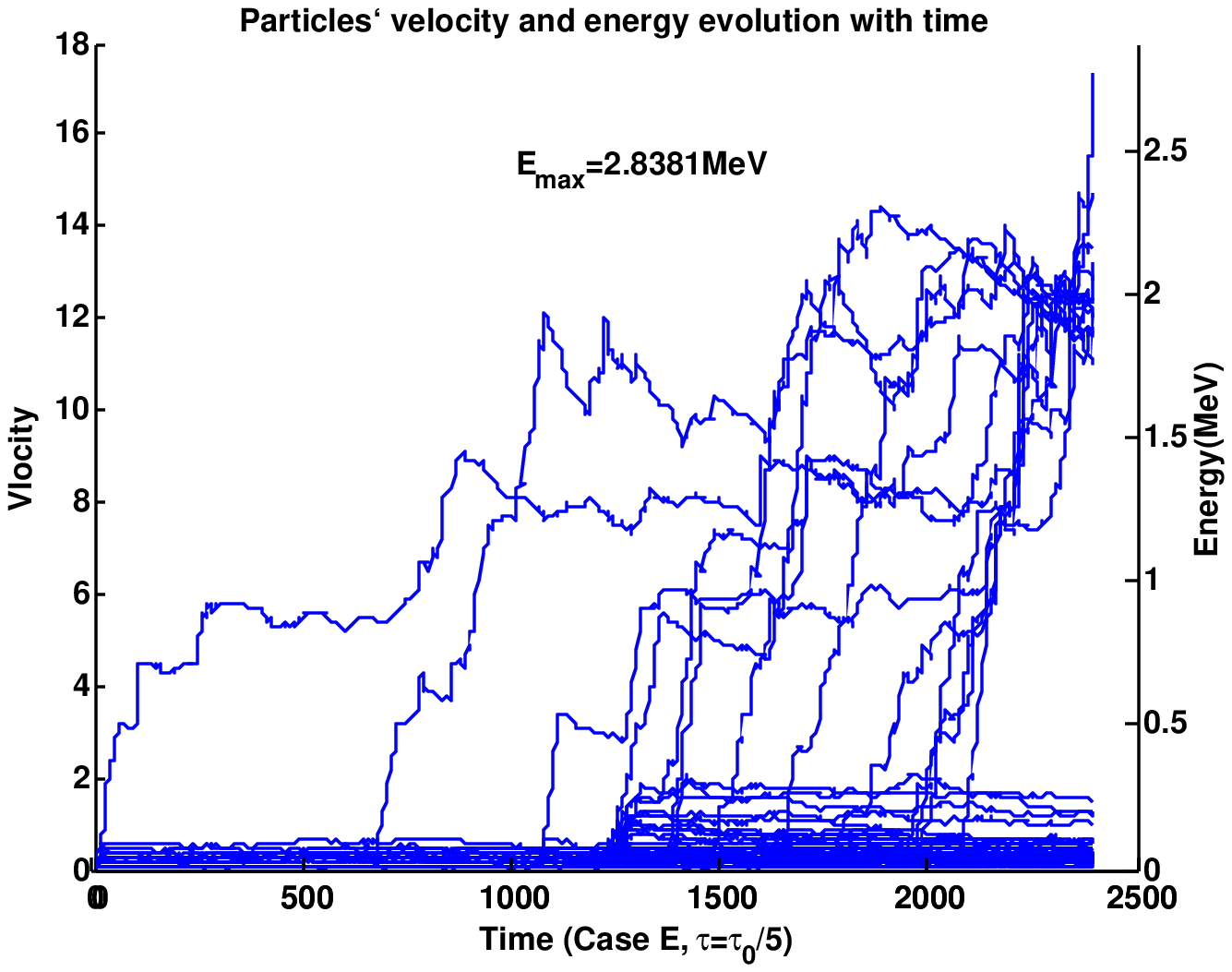}
  \includegraphics[width=2.8in, angle=0]{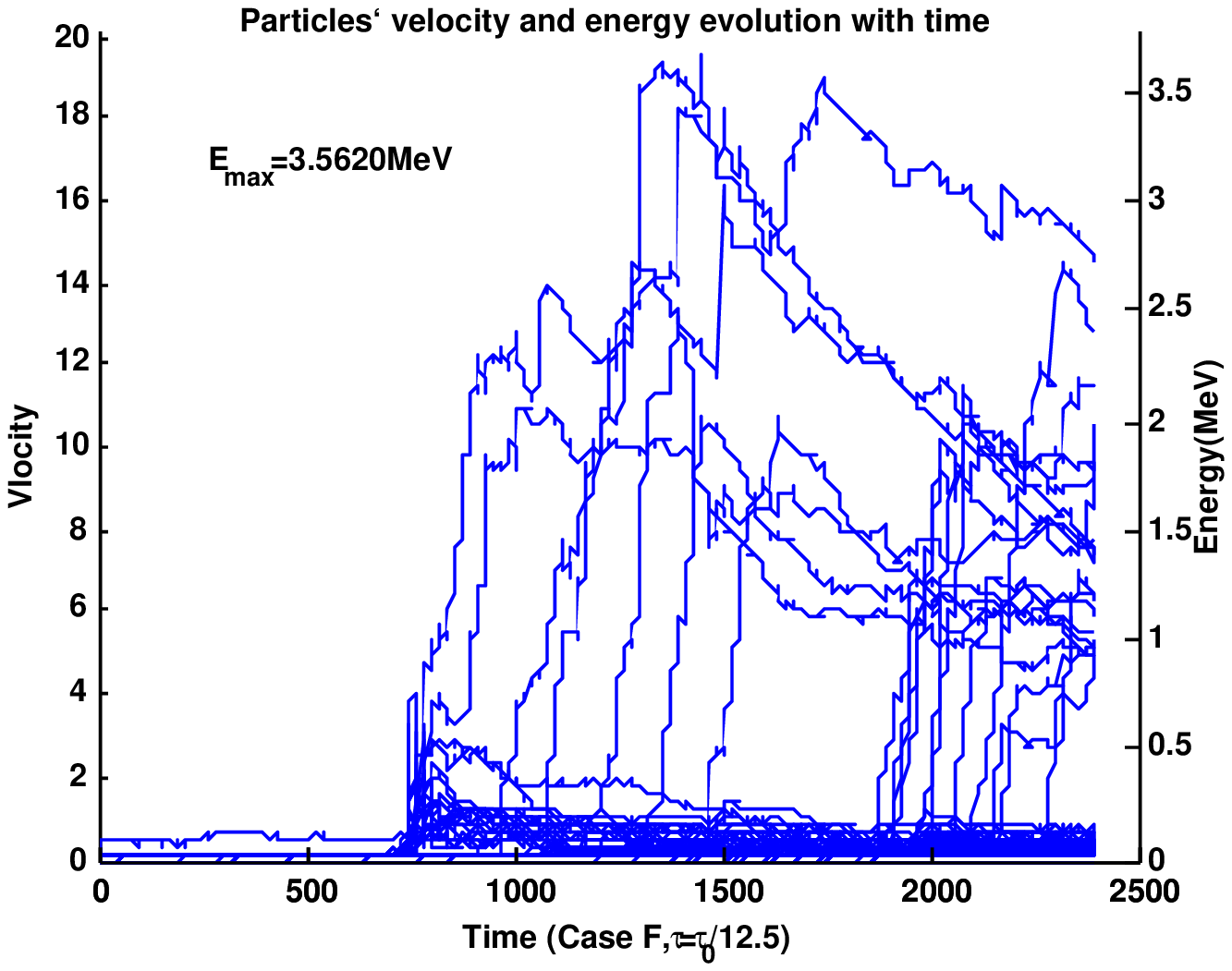}
  \caption{\tiny A number of particles are extracted from the simulation box in corresponding cases from Cases A,B,C,D,
  E, to Case F, respectively. Blue curve represents one particle's evolution of velocity and energy with time. The top peak
  in each plot shows the maximum velocity or energy in corresponding case. A part of particles have no acceleration
  at the bottom of each plot. A few curves with jumps
  from the lower energy to the higher energy in each plot indicate that they are accelerated in the shock with
  time. The $E_{max}$ in Case C with a scattering time
$\tau=\tau_{0}/3$ achieves an energy saturation at 5.5506MeV with comparing six cases.}
  \label{simp_fig1}
 \end{figure}
To inspect the entire particles' acceleration processes in the shock, we extracte a number of particles from
 the simulated box in each case. Six plots in Figure \ref{simp_fig1} are taken from six simulate cases from
 Cases A, B, C, D, E, to Case F, respectively. Each curve in each plot represents one particle's evolution of
 velocity and energy with time. Every plot have a few peak velocities in some accelerated particles, the highest peak value in the
 velocity or energy axis represents the maximum velocity or energy in corresponding case. Each maximum energy
 value is denoted in each plot. Among of these six cases, the maximum energy value in Case C achieves an
 energy saturation at 5.5506MeV. In addition, we can also find that a part of particles at the bottom of
 each plot exhibit no acceleration in total simulation time.  Another part of particles with jumps
 from the lower energy to the higher energy in each plot indicate their accelerating processes in the shock with
 time. Simultaneously, energy jumps in corresponding case show an increasing steeply tendency with a decreasing
 value for the constant of scattering time from Cases A, B, C, D, E, to Case F, respectively.
 Case F shows severely steep jumps and steeply descents in
 energy or velocity curves,because the scattering time $\tau=\tau_{0}/12.5$
 is chosen to approach to time step ($dt=\tau_{0}/25$). These results indicate that the computational
 accuracy requires the scattering time should be more enough than time
 step.

\subsection{The $E_{max}$ Function.}
Here, we focus on an isolated CME-driven shock for calculating the maximum particle energy E$_{max}$ in those cases
applying the different values for the scattering time. Using our dynamical Monte Carlo model, we have obtained the
different values for the E$_{max}$ in those cases. So we can build the function of the maximum particle energy
E$_{max}$ versus the values for the scattering time $\tau$ with the value from $\tau_{0}$, $\tau_{0}$/2, $\tau_{0}$/3,
$\tau_{0}$/4, and $\tau_{0}$/5 to $\tau_{0}$/12.5 in corresponding case.

\begin{figure}[t]
\centerline{
  \includegraphics[width=2.5in, angle=-90]{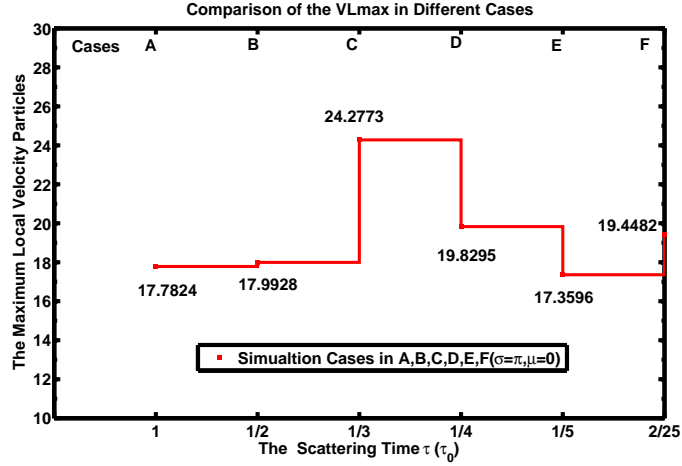}}
  \caption{The maximum local velocities $VL_{max}$ of the accelerated energetic particles
  in all cases are plotted
  as a stair function of the value for the scattering time. All these cases are simulated
  using Monte Carlo model complying with a Gaussian scattering angular
  distribution with a standard deviation value for
  $\sigma=\pi$, and average value for $\mu=0$.}
  \label{simp_fig2}
 \end{figure}

Utilizing the method described in Section \textbf{Method}, the calculations of $E_{max}$ are performed under the
scattering angular distribution with a standard deviation value for $\sigma=\pi$ and an average value for $\mu=0$,
which would relatively be more efficient for the particle acceleration in the CME-driven shock demonstrated by previous
model \citep{wy12}. From the large population of the accelerated particles at the end of the simulation in each case,
we find each maximum local velocity $VL_{max}$ in corresponding case with its scattering time $\tau$. The relationship
between the maximum local velocity $VL_{max}$ and the value for the scattering time $\tau$ in all cases are shown in
Fig.\ref{simp_fig2}. The solid line denotes the correlation of the maximum local velocity $VL_{max}$ versus the
different value for the scattering time $\tau$ in corresponding case. In present Monte Carlo model, the value for the
scattering time $\tau$ is chosen from $\tau_{0}$, $\tau_{0}$/2, $\tau_{0}$/3, $\tau_{0}$/4 and $\tau_{0}$/5 to
$\tau_{0}$/12.5, respectively.  And these six squares represent the maximum local velocity values for $VL_{max}$ in all
cases with their correspondingly values for the scattering time. The maximum local velocity $VL_{max}$ is represented
by a dimensionless value for $VL_{max(A)}$=17.7824, $VL_{max(B)}$=17.9928, $VL_{max(C)}$=24.2773,
$VL_{max(D)}$=19.8295, $VL_{max(E)}$=17.3596, and $VL_{max(F)}$=19.4482 in each case, respectively. As shown in
Fig.\ref{simp_fig2}, among these maximum local velocities $VL_{max}$, the largest is the one in Case C with a value of
$VL_{max(C)}$=24.2773, and its value for the scattering time is $\tau_{0}$/3. The top of the stair line in
Fig.\ref{simp_fig2} shows that there exists a saturation in a function of the maximum local velocity $VL_{max}$ versus
the value of the scattering time $\tau$ under the resonant diffusion scenarios.

\begin{figure}[t]
\centerline{
  \includegraphics[width=2.5in, angle=-90]{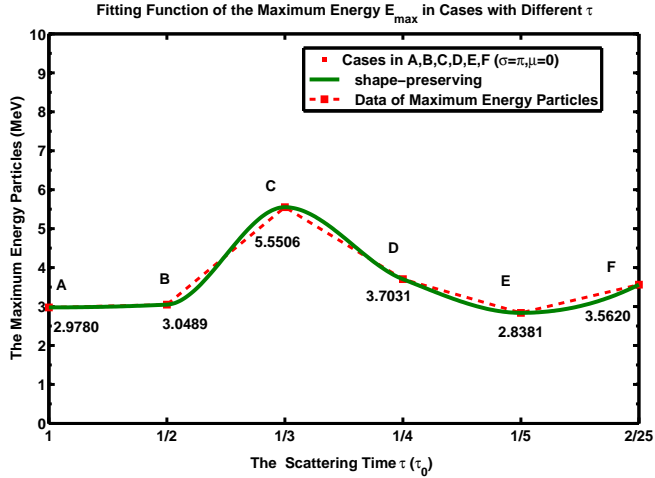}}
  \caption{The shape-preserving curve of the maximum particle energy $E_{max}$ as a function of
  the value for the scattering time in six cases. The values for maximum particle energy are indicated
  by scaled values along the profile of the shape-preserving curve in all cases with different value
  for the scattering time.
  The profile of the $E_{max}$ in different cases shows a saturation value for the
  maximum particle energy in Case C with a critical value for the scattering time
  $\tau_{0}/3$. But all of them are not exceed the upper limit of the $E_{break}$ at 10 MeV of observation.}
  \label{simp_fig3}
 \end{figure}

Fig.\ref{simp_fig3} shows the fitting curve of the maximum particle energy $E_{max}$ versus the values for the
scattering time $\tau$. The maximum particle energy $E_{max}$ are calculated in the shock frame by a scaled value
according to the scale factor for velocity $v_{scale}$. The maximum particle energy $E_{max}$ in each case varies along
the shape-preserving curve with a sequence of $E_{max(A)}$=2.9780MeV, $E_{max(B)}$=3.0489MeV, $E_{max(C)}$=5.5506MeV,
$E_{max(D)}$=3.7031MeV, $E_{max(E)}$=2.8381MeV,  and $E_{max(F)}$ =3.5620MeV  from Cases A, B, C, D, and E, to Case F.
respectively. All of those maximum particle energy $E_{max}$ are not exceed to the upper limit of E$_{break}$ at 10MeV
in the observation. But Case C with corresponding value for the scattering time $\tau_{0}/3$ shows that the largest
maximum particle energy is $E_{max(C)}$=5.5506MeV, which is still less than the upper limit of the E$_{break}$ region.
It implies that whatever the value for the scattering time is chosen under an isolated shock model, the maximum
particle energy $E_{max}$ would not more than the upper limit of the $E_{break}$ region in the observed energy
spectrum. According to these simulation results, the energy spectrum ``cut-off" would be formed near the energy range
of 5MeV. Simultaneously, the saturation value for the maximum energy function demonstrate that these maximum particle
energy $E_{max}$ can fit the observed lower energy spectrum below the $E_{break}$ limit. Looking from the
shape-preserving curve in Fig.\ref{simp_fig3}, the $E_{max}$ will not increase with the value for the scattering time
$\tau$ decreasing from $\tau_{0}$/3 to $\tau_{0}$/5. Although the function of maximum particle energy $E_{max}$ shows a
lightly ascendant tendency when the value for the scattering time decreasing from $\tau_{0}$/5 to $\tau_{0}$/12.5, the
value for the scattering time $\tau_{0}$/12.5 is approaching to the time step $dt$ (i.e., $dt=\tau_{0}/25$).
Considering the precision of the calculation, the value for the scattering time $\tau$ should be chosen to be not less
than the time step $dt$. Since the amplified magnetic field is limited by the order of the magnitude $\delta B/B\sim
1$, whatever the value for the scattering time is chosen in an isolated shock model, the obtained maximum particle
energy E$_{max}$  are not more than the upper limit of the E$_{break}$ in the observed energy spectrum. If we expected
to obtain a more extended energy spectrum beyond the upper limit of $E_{break}$ at 10MeV and even up to GeV, the
multiple shocks model would be applied. So it means that the efficiency of acceleration in an isolated shock model will
not exceed the upper limit of the $E_{break}$ as long as the value for the scattering time is chosen to be enough more
than the time step $dt$. In addition, it also implies that the realistic observation of the $E_{break}$ energy spectrum
strongly requires a multiple shock model to transfer the shock's energy into the superthermal particles up to a highest
energy spectrum for explaining the $E_{break}$ formation and the higher energy spectrum.

\subsection{The Energy Spectra}
\begin{figure}[t]
\centerline{
  \includegraphics[width=4.0in,angle=-90]{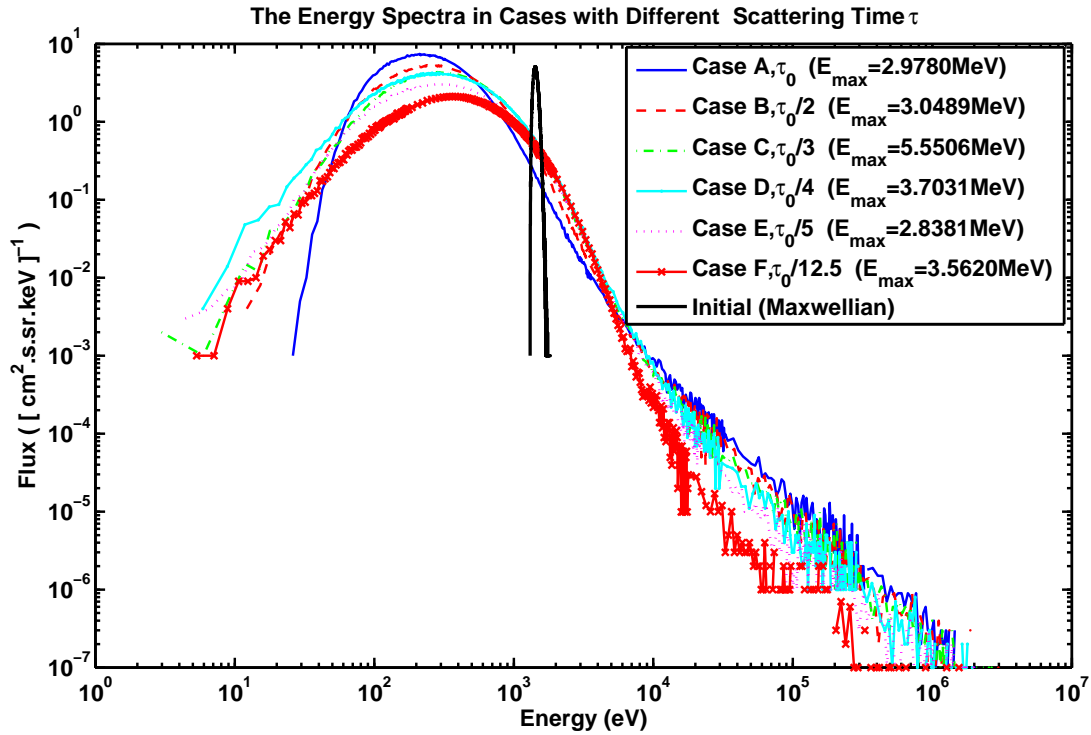}}
  \caption{The energy spectra obtained from downstream region in six cases with different
   values for the scattering
   time. The thick solid line with a narrow peak at $E =
   $1.4315keV represents the initial Maxwellian energy distribution in the
   upstream region.
   All of cases consist with the low energy spectrum with all $E_{max}$ less than 10MeV.}
  \label{double_fig4}
 \end{figure}

Fig.\ref{double_fig4} shows the shock energy spectra calculated in the downstream region in all cases. As far as the
shape of the energy spectrum is concerned, the power-law slope of six extended curves are similar, because all cases
are done in the same resonant diffusion scenarios just only with different values for the scattering time. However,
among these cases, the energy spectrum in Case C with the value for the scattering time of $\tau_{0}$/3 shows a
relatively hard slope in the highest energy spectral tail. Under an isolated shock model, each case shows that how the
initial Maxwellian energy spectrum to evolve into the extended energy spectrum with ``power-law" structure in its high
energy, respectively. By comparison, we calculated the average value of the maximum particle energy in present six
cases. The average value for maximum particle energy is $<E_{max}>$=3.6135MeV and the average value for energy spectral
index is $\Gamma\sim1.125$. These results agree with the low energy spectrum in the observations from the multiple
spacecraft. Observed energy spectrum\citep{mewaldt08} shows low energy spectrum with an index of $\Gamma=1.07$ and a
high energy spectrum with an index of $\Gamma=2.45$. The observed energy spectrum indicated that there exists an
E$_{break}$ between the lower energy spectrum and the higher energy spectrum. From these simulated cases, we concluded
that all these energy spectra are characterized by a ``power-law" with an averaged index $\Gamma\sim1.125$, which
consists with the observed index $\Gamma=1.07$ of the low energy spectrum. Since there is no maximum particle energy
$E_{max}$ in these six cases beyond the upper limit of $E_{break}$ at 10MeV, we are unable to conclude that there
should exist an E$_{break}$ at 1-10MeV as a ``break" from the lower energy spectrum to higher energy spectrum at this
range. If we expect the second higher energy spectrum to occur, we can guess that there must exist an enhancement in
amplification of the magnetic field associated with multiple shocks model. We propose the multiple shocks model would
be applied to further investigate the higher energy spectrum in CME shock events. Recently, there are some analysis of
multiple CME collision events that have been discussed. For example, \citet{czd13} report the initiation process of
compound CME activity consisting of two successive eruptions of flare ropes that occurred on January 23 2012. Another
example presented by \citet{ll14} shows that the interactions between consecutive CMEs resulted in a ``perfect storm"
near 1 AU on 23 July 2012, which would induce to nonlinear amplification of magnetic field. Further more evidences
could be gathered from observations in spacecraft such as SDO, SOHO, ACE, Wind and etc. In the implication from these
present simulated results, we propose to build a multiple shocks model to simulate the E$_{break}$ formation and the
higher energy spectrum in the interplanetary shock. In present model, we think that the parameter of the scattering
time would play key role on the strength of the diffusive coefficient for E$_{max}$ production within the resonant
diffusion scenarios associated with isolated shock. According to the final results, we find the relationship between
the maximum particle energy E$_{max}$ and the different value for the scattering time in isolated shock model. Although
the difference on these maximum particle energy E$_{max}$ in simulated cases has happened, no maximum particle energy
E$_{max}$ can exceed the upper limit of $E_{break}$ to further evolve into a higher energy spectrum up to GeV. The
future simulation is necessary to verify the higher energy spectrum with an index of $\Gamma\simeq2.5$ and the energy
spectral ``break" formation by applying multiple shocks model.


\section{Summary and conclusions}\label{sec-summary}
In summary, these presented simulations are unable to verify that there should exist an energy spectral ``break" below
10MeV in some large CME-driven shocks. But instead we verify that there is an energy spectral ``cut-off" near the range
of energy at 5MeV in an isolated CME-driven shock. We calculate the maximum particle energy E$_{max}$ focusing on Dec
14 2006 CME-driven shock event and built the relationship between the maximum particle energy E$_{max}$ and the value
for the scattering time $\tau$. We find that the maximum particle energy E$_{max}$ approaches to a saturation near 5MeV
below the upper limit of $E_{break}$ of the observed energy spectrum. We verify the lower energy spectrum is consistent
with the observed low energy spectrum, but no higher energy spectrum appears. Although there are several large SEP
events in the past solar cycle 23 appear the energy spectral ``breaks" between 1-10MeV, there is still no very
reasonable explanation. Since these observations are depend on multiple spacecraft, it is not easy to treat the system
errors and couple the observed data obtained from different spatial orientations. The huge computational expense also
limits numerical methods to reach an enough high energy spectral tail for further to identify this ``break". In the
view of the current theoretical point about DSA, analytic method give an implication that this ``break" would be
connected with particle leakage mechanism. This ``break" seemly can be predicted in the escaped position ahead of
supernova remnants (SNRs) shock, where the SNRs has a collision with nearby molecule clouds (MC). This idea will
motivate us to further investigate the energy spectrum $E_{break}$ formation and the higher energy spectrum. Hopefully,
we propose the multiple shocks model would play key role on explaining the energy spectrum $E_{break}$ formation and
the higher energy spectral shape.

\begin{acknowledgements}
{Present work is supported by Xinjiang Natural Science Foundation No. 2014211A069. This work is also funded by the Key
Laboratory of Solar Activities of NAOC, the Key Laboratory of Modern Astronomy and Astrophysics (Nanjing University),
Ministry of Education, and China Scholarship Council(CSC). Simultaneously, authors thank Profs. Joe Giacalone in
University of Arizona and H. B. Hu in Institute of High Energy Physics of Chinese Academy of Sciences for their very
helpful comments and discussions on this paper.}
\end{acknowledgements}

\end{document}